# Closed-Form Path-Loss Predictor for Gaussianly Distributed Nodes

Mouhamed Abdulla and Yousef R. Shayan
Department of Electrical and Computer Engineering
Concordia University
Montréal, Québec, Canada
Email: {m_abdull, yshayan}@ece.concordia.ca

***Abstract*—The emulation of wireless nodes spatial position is a practice used by deployment engineers and network planners to analyze the characteristics of a network. In particular, nodes geolocation will directly impact factors such as connectivity, signals fidelity, and service quality. In literature, in addition to typical homogenous scattering, normal distribution is frequently used to model mobiles concentration in a cellular system. Moreover, Gaussian dropping is often considered as an effective placement method for airborne sensor deployment. Despite the practicality of this model, getting the network channel loss distribution still relies on exhaustive Monte Carlo simulation. In this paper, we argue the need for this inefficient approach and hence derived a generic and exact closed-form expression for the path-loss distribution density between a base-station and a network of nodes. Simulation was used to reaffirm the validity of the theoretical analysis using values from the new IEEE 802.20 standard.**

## I. Introduction

In trying to understand the behavior of medium and large networks for purpose of analysis, engineers always attempt to emulate what is most likely occurring in real-life scenarios using stochastic models. In wireless communications, the spatial distribution of wireless nodes is a topic that directly affects all lower layers of the OSI model: from the physical to the network. Though, more specifically, users' density will impact the network capacity, the coverage area, the connectivity, the power consumption, and interference, among others. Thus, such topic should equally be of interest to network designers and deployment engineers dealing with mobiles or Wireless Sensor Networks (WSN).

In fact, in order to get near real time sense of nodes density, technology based methods are perhaps the best way to provide this sort of observation. The most popular among them is the Global Positioning System (GPS) [1]. However, the accuracy of GPS, at least for civilian usage is still not to the level required for high-precision positioning. Though, other standards with better fidelity may also be used to gain location information such as Bluetooth [2]. Yet, despite the technology selected, the added overhead and hardware complexity to nodes will introduce newer design challenges in addition to manufacturing cost.

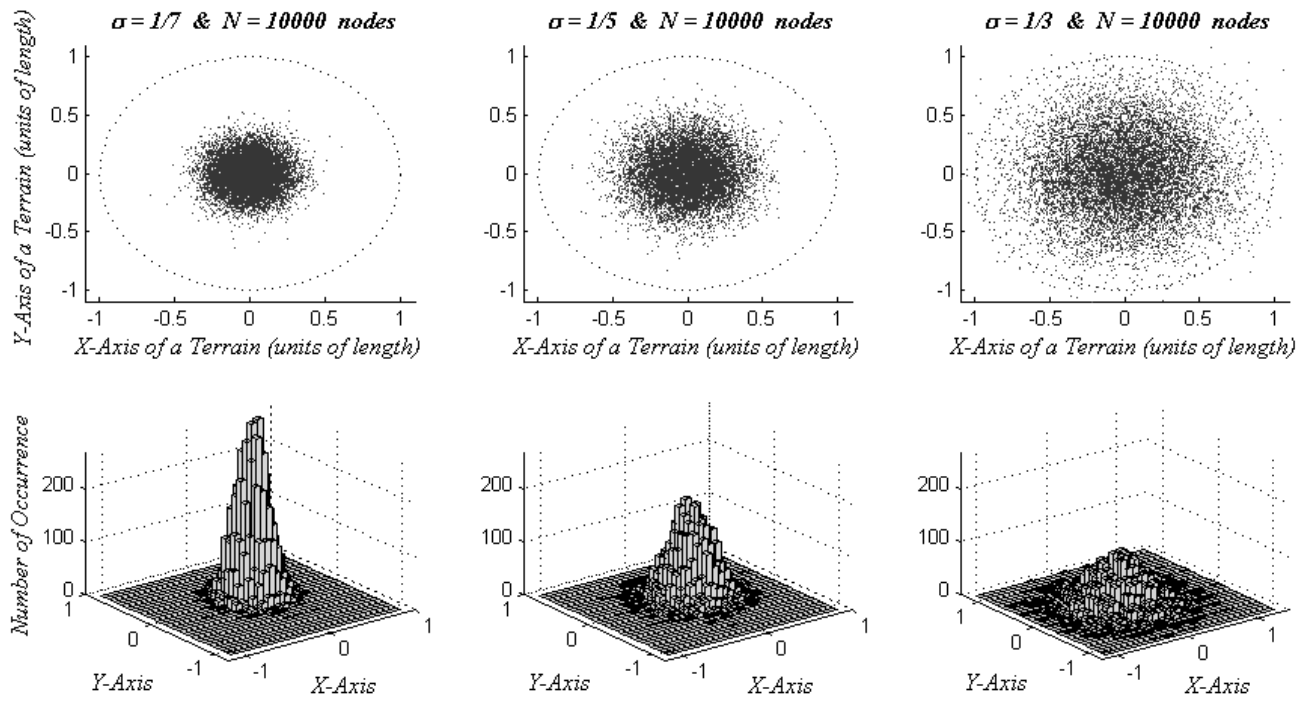

Figure 1. Normally distributed mobiles with different spreads.

Apart from this method, human behavior through social patterns and inclinations is another interesting alternative for spatial prediction [2]. In general, behavior is driven by physical conditions, emotional states, cognitive capabilities and social status [3]. Clearly, this is only an elementary model; fully understanding the human behavior is still far from being properly solved or understood. Moreover, social trends, for the most part, require a lot of effort and resources due to the need for long and intense observations. And the spatial densities obtained through them are not necessarily universal or generic for diverse network projects, and are hence useful only for the observed site.

Indeed, certain papers have assumed that terminals position is deterministic; in other words the location of nodes are known in advance [4] [5]. Yet, non-random or predetermined system placement is only feasible for fixed antennas such as static sensor networks, and is unpractical for quasi-stationary nodes and fully mobile devices.

Sometimes, Probability Density Functions (PDFs) for users' position traffic are conjectured based on assumptions [6] [7]. Though, the most common hypothesis of nodes distribution used among researchers is founded on the uniform density [8]–[16]. Admittedly, uniform distribution is realistic only if a large surface area has no natural or manmade topographical features, because users by their nature tend to cluster [17] [18]. Needless to say, nearly most of earth surface, if not all, contains characteristic features such as rivers, mountains, irregular grounds, buildings, diverse infrastructures, etc. Thus, homogenous distribution is merely a quick oversimplification of the matter. Nonetheless, uniform density remains reasonable and to some extent practical in simulations because of its associated simplicity.

Besides, some researchers have even considered dropping nodes in a Gaussian fashion in order to effectively emulate point process statistics with denser clustering near the Base Station (BS) [10], [16], [18], [19]–[23]. It was also reported that a Gaussian distribution is an effective way to model nodes deployment from an air moving vehicle such as an airplane or a rotorcraft, with practicality in rural area monitoring or hostile environments [24]. In other words, if a set of sensors are intended to be positioned about a specific location *P* (say a pre-deployed access point), once dropped, they are expected to be anywhere in a cloud around *P* due to factors such as wind, speed, height, etc [25]. Hence, based on the central limit theorem the sensor constellation will follow a normal PDF [26].

As it can be noticed form the above survey, the two commonly employed, cost-effective, stochastic postulations for analysis are based on uniform and Gaussian spatial methods. In fact, in comparison, the normal model gives at least a degree of freedom to network designer for controlling nodes scattering through the standard deviation variable of the joint probability: $\sigma \in \mathbb{R}^+$ . Fig. 1 shows an example of random dropping for a fix amount of nodes over diverse spreads.

Moreover, during network analysis, wireless channel corruption such as Path-Loss (PL) is always vital and critical. Thus, being able to predict the PL by its associated density becomes very useful. By and large, the only way to obtain an estimate of the PL distribution for a particular network relies on computationally complex Monte Carlo simulation for each network being researched. Indeed, perhaps due to its inherent simplicity, the PDF of the channel loss for the homogenous case has already been obtained [8]. However, there is yet a reporting to be made for the Gaussian assumption. As a result, given the wide practicality of the Gaussian model, in this contribution we will derive a generic and exact closed-form distribution expression for the PL between an Access Point (AP) and a node.

## II. Random Generation

The aim of this paper is to find the channel loss density of normally dropped nodes. Thus we start form the 2D Gaussian distribution given in (1), where $x_G = (x - m_X)/\sigma_X$ and $y_G = (y - m_Y)/\sigma_Y$ such that $\sigma$ and $m$ are the spread and mean along each axis, and $\rho_{XY}$ is the correlation coefficient.

$$f_{XY}(x, y) = \frac{\exp\left\{-\left(x_G^2 + y_G^2 - 2\rho_{XY} x_G y_G\right) \big/ 2\left(1 - \rho_{XY}^2\right)\right\}}{2\pi\sigma_X \sigma_Y \sqrt{1 - \rho_{XY}^2}} \tag{1}$$

In the above expression, we could immediately set the means to zero since the BS or AP will be at the origin. Also, past contributions have always assumed the spatial axes to be uncorrelated. In fact, this is done not only for the sake of simplicity, but also from an intuitive perspective because no *a priori* statistical knowledge of users' trends or terrain limitations is known. Further, in the generation in order to have a single controllable agent, we will assume the Standard Deviation (SD) spread along each axis to be the same.

As for stochastic generation, most high-level computer languages have a built-in Pseudo-Random (PR) sequence. And the numbers are usually generated by $U(0,1)$ , where $U(a,b) = 1/(b-a)$ is a uniform distribution for $x \in (a,b)$. Also, these values are obtained either through the *multiplicative congruential algorithm* or *Marsaglia's generator*. MATLAB, uses the later with some modification to produce a very long PR sequence of length $2^{1492} \approx 1.37 \times 10^{449}$ [27]. Moreover, several software programs, including MATLAB, are capable, based on the *ziggurat algorithm*, to generate values from the standard normal distribution. Notably, for PL analysis in a downlink/uplink scheme, we will need the random generation of the separation between a node and an AP, which is given by $R = \sqrt{X^2 + Y^2}$ where $X$ and $Y$ are each generated form a zero-mean Gaussian PDF.

If we assume a constant amount of random nodes, say 10,000, then as evident from Fig. 2 samples from a histogram properly overlap the theoretical curves for *x*, *y*, and *r* under different SDs. This further justifies the generation in addition to the spatial interpretation of Fig. 1. More specifically, this verification was needed so as to effectively be used as a benchmark for comparing our anticipated closed-form expression to Monte Carlo simulated samples.

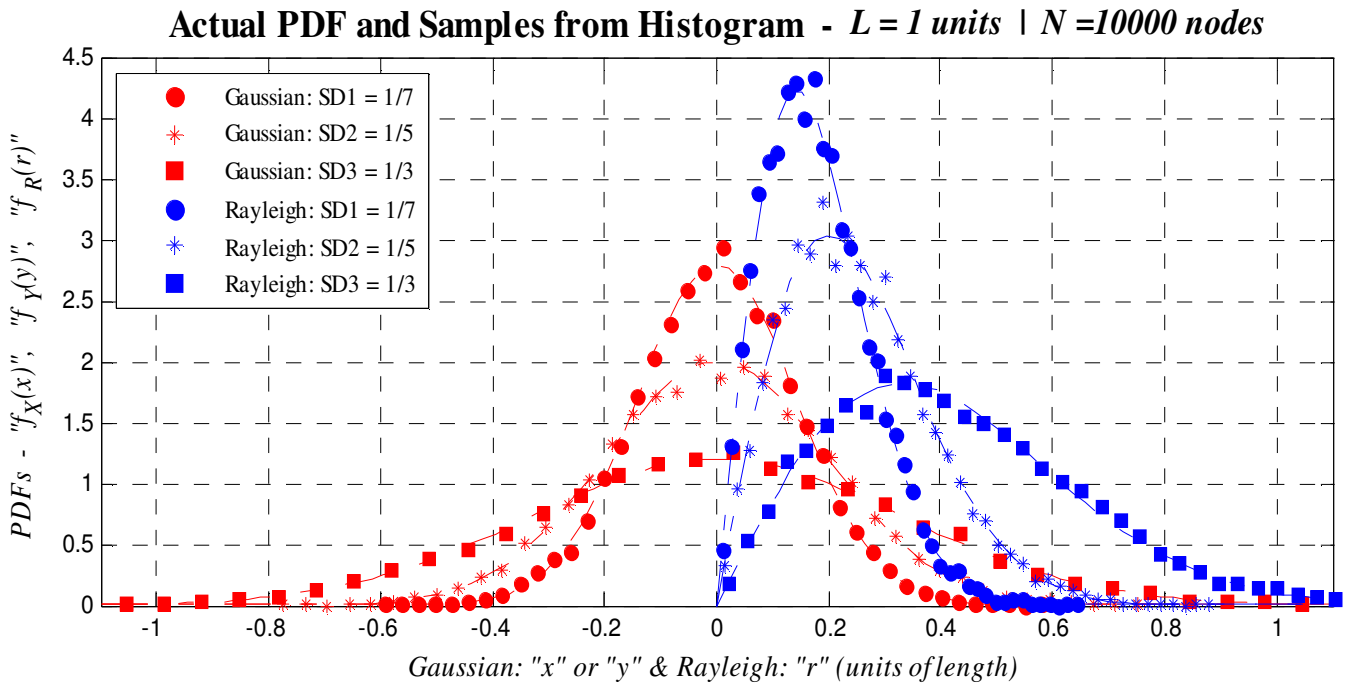

Figure 2. Theoretical and generated PDFs over different spreads.

## III. Path-Loss Distribution Model

In wireless communications, losses in the channel are an important contributor for signal corruption. Most notably the main losses in transmission are caused by path, shadowing, and fading, with probing accuracy of $\Delta \approx$ *1000λ*, *40λ*, and *λ* meters respectively; where *λ* is the wavelength of the carrier frequency. To only take into account the effect of large-scale attenuation we will base our analytical derivation on path-loss and shadowing, and ignore small-scale multipath fluctuations.

Admittedly, in most wireless applications, the power received or SNR is fixed at a certain threshold. On the other hand, the power transmitted by a telecommunication device or sensor is random due to the stochastic nature of the PL. Therefore, for any wireless application or system, it is very critical and useful to predict the behavior of the PL through a PDF. More specifically, it is important in power control of cellular systems for adequate handling of the near-far effect and inter-cell interference [28]. In fact, obtaining the PL density will give insight into power consumption, coverage, detection, sensing capability, etc [29].

There are various forms for the PL, though the most universal and widely accepted model has distance dependency as given by:

$$\begin{aligned} PL(r)_{dB} &= \overline{PL(r)}_{dB} + \Psi \\ &= \overline{PL(r_0)}_{dB} + 10n\log_{10}\left(\frac{r}{r_0}\right) + \Psi \end{aligned} \tag{2}$$

where $\overline{PL(r)}_{dB}$ is the average PL in decibels (dB), $\Psi$ is a Random Variable (RV) measured in dB representing the effect of shadowing with a log-normal distribution, i.e. $N(0,\sigma_\Psi^2)$, such that $N(m,\sigma^2)$ is a general Gaussian curve, and $\sigma_\Psi$ is the SD for shadowing also in dB. Further, $n$ is the PL exponent which depends on the propagation environment such as the existence of Line of Sight (LOS) or otherwise. Also, $r_0$ is referred to as the close-in distance measured in meters, and $r \geq r_0$ is the separation distance between a transmitter and a receiver. It is worth adding that the average PL at the close-in distance can be obtained empirically or for the simplest case, through the use of the Friis free space model. For ease of mathematical manipulations, we may represent (2) by mapping it to:

$$L_P = \overbrace{\alpha + \beta\log_{10}\left(\frac{r}{r_0}\right)}^{Path-Loss} + \overbrace{\Psi}^{Shadowing} = W + \Psi \tag{3}$$

First, we need to find the density of $r$ which is the distance between a random node and the corresponding BS or AP. To achieve this, it becomes natural to transform the PDF of (1) to polar notation because we want to find the distribution for the radius, where $J(r,\theta)$ is a 2D Jacobian matrix. Then, we determine the marginal density along $r \in \mathbb{R}^+$ as given by:

$$\begin{aligned} f_R(r) &= \int_{\theta=0}^{2\pi} \left[f_{XY}(x,y)\right]_{\substack{x=r\cos(\theta)\\ y=r\sin(\theta)}} \left|J(r,\theta)\right| d\theta \\ &= \frac{r\exp\left\{-r^2/2\sigma^2\right\}}{\sigma^2} \end{aligned} \tag{4}$$

Next, if we go back to (3) and focus on obtaining a distribution for the PL component only, while remembering that $\alpha$ and $\beta$ are deterministic scalars, we get for $w \in \mathbb{R}$:

$$f_W(w) = \frac{r_0 10^{(w-\alpha)/\beta}\ln(10)}{\beta}\left[f_R(r)\right]_{r=r_0 10^{(w-\alpha)/\beta}} \tag{5}$$

$$f_W(w) = \frac{r_0^2 10^{2(w-\alpha)/\beta}\ln(10)}{\beta\sigma^2}\exp\left\{-r_0^2 10^{2(w-\alpha)/\beta}/2\sigma^2\right\} \tag{6}$$

Pursuing this further, it should be stressed that both RVs $W$ and $\Psi$ for PL and shadowing are statistically independent, i.e. $\mathrm{E}[W\Psi] = \mathrm{E}[W]\mathrm{E}[\Psi]$. Therefore, to obtain the PDF of $L_P$ we would need to convolve the density of both terms as shown by:

$$\begin{aligned} f_{L_P}(l) &= f_W(l) * f_\Psi(l) \\ &= \int_{-\infty}^{\infty} f_W(\tau) f_\Psi(l-\tau)\, d\tau \triangleq \int_{-\infty}^{\infty} f(\tau)\, d\tau \qquad l \in \mathbb{R} \end{aligned} \tag{7}$$

where the first component inside the integral can be obtained from (6) after a swap of $w$ to the dummy variable $\tau$; as for the shadowing part, it is given by:

$$f_\Psi(l-\tau) = \frac{1}{\sqrt{2\pi}\sigma_\Psi}\exp\left\{-(\tau-l)^2/2\sigma_\Psi^2\right\} \quad \tau \in \mathbb{R} \tag{8}$$

All together, the integrand of (7) becomes:

$$\begin{aligned} f(\tau) = &\left\{\frac{r_0^2\ln(10)10^{-2\alpha/\beta}}{\sqrt{2\pi}\beta\sigma^2\sigma_\Psi}\right\} \overbrace{10^{2\tau/\beta}\exp\left\{-(\tau-l)^2/2\sigma_\Psi^2\right\}}^{\triangleq f_1(\tau)} \\ &\times \exp\left\{-\left(r_0^2 10^{-2\alpha/\beta}/2\sigma^2\right)10^{2\tau/\beta}\right\} \end{aligned} \tag{9}$$

After several mathematical manipulations and labor the exponential parts (i.e. "*e*" and "*10*") of $f_1(\tau)$ could be modified to have the $\tau$ entities combined together:

$$\begin{aligned} f_1(\tau) = &\exp\left\{2\ln(10)\left(\ln(10)\sigma_\Psi^2 + \beta l\right)/\beta^2\right\} \\ &\times \exp\left\{-\left(\tau - \left\{l + 2\ln(10)\sigma_\Psi^2/\beta\right\}\right)^2/2\sigma_\Psi^2\right\} \end{aligned} \tag{10}$$

Now, we substitute (10) into (9); then the result is used in the expression of (7). Further, to simplify the integrand, we perform the following transformation: $u = \tau - m$ where $m = l + 2\ln(10)\sigma_\Psi^2/\beta$. At this point, after carrying the convolution we obtain the *exact closed-form* PL density as shown below, where $l \in \mathbb{R}$:

$$\begin{aligned} f_{L_P}(l) = &\left\{\frac{r_0^2\ln(10)}{\sqrt{2\pi}\beta\sigma^2\sigma_\Psi}\right\}10^{2(l-\alpha)/\beta}\exp\left\{\left(\sqrt{2}\sigma_\Psi\ln(10)/\beta\right)^2\right\} \\ &\times \int_{u=-\infty}^{\infty}\exp\left\{-\frac{u^2}{2\sigma_\Psi^2} - \frac{r_0^2}{2\sigma^2}10^{2\left\{\beta(l-\alpha)+2\ln(10)\sigma_\Psi^2+\beta u\right\}/\beta^2}\right\}du \end{aligned} \tag{11}$$

In (11) the integral part can be evaluated using any numerical integration method such as Simpson's or trapezoidal rule. Though, we could continue the derivation and attempt to express the result in series notation. In principle, notice that the integration is of the form:

$$I_\infty \triangleq \int_{x=-\infty}^{\infty} \exp\left(-ax^2 - b\cdot 10^{cx}\right)dx \qquad (12)$$

If we expand the base-10 part using Taylor series we converge to the expression of (13), where a sample of the associated $\boldsymbol{\pi}_{(i,j)} \in \mathbb{N}^*$ are given in Table I.

$$\exp\left(-b\cdot 10^{cx}\right) = \sum_{i=0}^{\infty} \frac{x^i}{i!}\left\{ d^i\left(e^{-b\cdot 10^{cx}}\right)\Big/dx^i \right\}\Bigg|_{x=0}$$
$$= e^{-b}\left\{ 1 + \sum_{i=1}^{\infty}\sum_{j=0}^{i-1} \frac{(-b)^{j+1}}{i!}\left(c\ln(10)\right)^i \boldsymbol{\pi}_{(i,j)} x^i \right\} \qquad (13)$$

TABLE I. ANALYTICAL VALUES FOR THE SERIES OF EQUATION (13)

| $\boldsymbol{\pi}_{(i,j)}$ | $j=0$ | $j=1$ | $j=2$ | $j=3$ | $j=4$ |
|---|---|---|---|---|---|
| $i=1$ | 1 | $\cdots$ | $\cdots$ | $\cdots$ | $\cdots$ |
| $i=2$ | 1 | 1 | $\cdots$ | $\cdots$ | $\cdots$ |
| $i=3$ | 1 | 3 | 1 | $\cdots$ | $\cdots$ |
| $i=4$ | 1 | 7 | 6 | 1 | $\cdots$ |
| $i=5$ | 1 | 15 | 25 | 10 | 1 |

Now, we bring (13) into (12) to obtain:

$$I_\infty = e^{-b}\left\{ \begin{array}{l} \displaystyle\int_{x=-\infty}^{\infty} \exp\left(-ax^2\right)dx + \sum_{i=1}^{\infty}\sum_{j=0}^{i-1}\frac{(-b)^{j+1}}{i!}\left(c\ln(10)\right)^i \boldsymbol{\pi}_{(i,j)} \\ \times \underbrace{\displaystyle\int_{x=-\infty}^{\infty} \exp\left(-ax^2\right)x^i dx}_{\triangleq \Lambda_i} \end{array} \right\} \qquad (14)$$

At his level, we notice that the integration in (14) can be represented by the general class of the Gaussian integral, in other words:

$$\Lambda_i = \int_{x=-\infty}^{\infty} \exp\left(-ax^2\right)x^i dx = \begin{cases} 2I_i(a) & i = 0,2,4,\cdots \\ 0 & i = 1,3,5,\cdots \end{cases} \qquad (15)$$

where $I_i(a)$ is given in (16) [30], and the definition of the double factorial is available in [31]. In fact, after careful scrutiny, we found a more compact representation for *n!!* as represented in (17).

$$I_i(a) = \int_{x=0}^{\infty} \exp\left(-ax^2\right)x^i dx = \begin{cases} \dfrac{(i-1)!!}{2^{(i/2+1)}a^{i/2}}\sqrt{\dfrac{\pi}{a}} & i = 0,2,4,\cdots \\ \dfrac{\left((i-1)/2\right)!}{2a^{(i+1)/2}} & i = 1,3,5,\cdots \end{cases} \qquad (16)$$

$$n!! = \begin{cases} 1 & n = -1,0 \\ 1.3.5\cdots(n-2).n & n = 1,3,5,\cdots \\ 2.4.6\cdots(n-2).n & n = 2,4,6,\cdots \end{cases} = \begin{cases} \dfrac{(n+1)!}{2^{(n+1)/2}\left((n+1)/2\right)!} & n = -1,1,3,5,\cdots \\ 2^{n/2}(n/2)! & n = 0,2,4,6,\cdots \end{cases} \qquad (17)$$

At present, once we substitute (17) into (16) we obtain:

$$I_i(a) = \int_{x=0}^{\infty} \exp\left(-ax^2\right)x^i dx = \begin{cases} \dfrac{i!}{2^{(i+1)}(i/2)!}\sqrt{\dfrac{\pi}{a^{(i+1)}}} & i = 0,2,4,\cdots \\ \dfrac{\left((i-1)/2\right)!}{2a^{(i+1)/2}} & i = 1,3,5,\cdots \end{cases} \qquad (18)$$

Using the finding of (18), we simply (15):

$$\Lambda_i = \begin{cases} \dfrac{i!}{2^i(i/2)!}\sqrt{\dfrac{\pi}{a^{(i+1)}}} & i = 0,2,4,\cdots \\ 0 & i = 1,3,5,\cdots \end{cases} \qquad (19)$$

At this point we plug (19) into (14):

$$I_\infty = e^{-b}\sqrt{\frac{\pi}{a}}\left\{ 1 + \sum_{i=2,4,6\ldots}^{\infty}\sum_{j=0}^{i-1}\frac{(-b)^{j+1}}{(i/2)!}\left(\frac{c\ln(10)}{2\sqrt{a}}\right)^i \boldsymbol{\pi}_{(i,j)} \right\} \qquad (20)$$

Finally, we perform a change in the index of the inner summation, that is we set $i = 2k$ for $k \in \mathbb{N}^*$. Thus, the final result for $l \in \mathbb{R}$ becomes:

$$f_{L_p}(l) = \left\{\frac{r_0^2\ln(10)}{\beta\sigma^2}\right\}10^{\frac{2(l-\alpha)}{\beta}} \exp\left\{ \left(\frac{\sqrt{2}\sigma_\Psi \ln(10)}{\beta}\right)^2 - \frac{r_0^2\cdot 10^{\frac{2\{\beta(l-\alpha)+2\ln(10)\sigma_\Psi^2\}}{\beta^2}}}{2\sigma^2} \right\}$$
$$\times\left\{ 1 + \sum_{k=1}^{\infty}\sum_{j=0}^{2k-1}\frac{\boldsymbol{\pi}_{(2k,j)}}{k!}\left(\frac{\sqrt{2}\ln(10)\sigma_\Psi}{\beta}\right)^{2k}\left(\frac{-r_0^2\cdot 10^{2\{\beta(l-\alpha)+2\ln(10)\sigma_\Psi^2\}/\beta^2}}{2\sigma^2}\right)^{j+1} \right\} \qquad (21)$$

## IV. Simulation Results

For simulation, we will utilize parameters from the Mobile Broadband Wireless Access (MBWA) standard [32]. The values used for different channel environments are shown in Table II, where the carrier frequency assumed is 1.9 GHz. Also, in the table, note that the close-in distance is not explicitly given, and is in fact absorbed by "$\alpha$" of the PL.

The simulation is based on 10,000 random samples using the Gaussian profile. Then, the PL is measured for each generated position. Following this, a PL histogram for the entire network is obtained. Next, we scale the histogram to obtain the PDF equivalent so as to be compared to the exact closed-form expression of the previous section. We also obtained the respective Cumulative Distribution Function (CDF) for all cases. This is done because the CDF is an appropriate delivery metric for the network Quality of Service (QoS). In other words, the network constellation with the smallest overall PL at CDF saturation guarantees signals with higher fidelity.

TABLE II. IEEE 802.20 CHANNEL MODELS

| **Channel Environment** | *Suburban Macrocell* | *Urban Macrocell* | *Urban Microcell* | |
|---|---|---|---|---|
| **Cell Radius** [km] | $0.6 \leq L \leq 3.5$ | $0.6 \leq L \leq 3.5$ | $0.2 \leq L \leq 0.3$ | |
| **Propagation Model** | *COST-231 Hata-Model* | *COST-231 Hata-Model* | *COST-231 Walfish-Ikegami* | |
| **Standard Deviation for Shadowing** [dB] | 10 | 10 | NLOS: 10 | LOS: 4 |
| **Path-Loss** [dB] | $\alpha = 31.5$, $\beta = 35$ | $\alpha = 34.5$, $\beta = 35$ | $\alpha = 34.53$, $\beta = 38$ | $\alpha = 30.18$, $\beta = 26$ |
| **Supported Distances** [m] | $r_0 = 35 \leq r \leq L$ | $r_0 = 35 \leq r \leq L$ | $r_0 = 20 \leq r \leq L$ | |
| **Mobility** [km/hr] | $0 \rightarrow 250$ | $0 \rightarrow 250$ | $0 \rightarrow 120$ | |

Fig. 3 shows the simulation for diverse spatial spreads, links, and channel parameters. As it can be observed, the values obtained from Monte Carlo simulation properly match the theoretical closed-form expression derived in this treatment over the various cases.

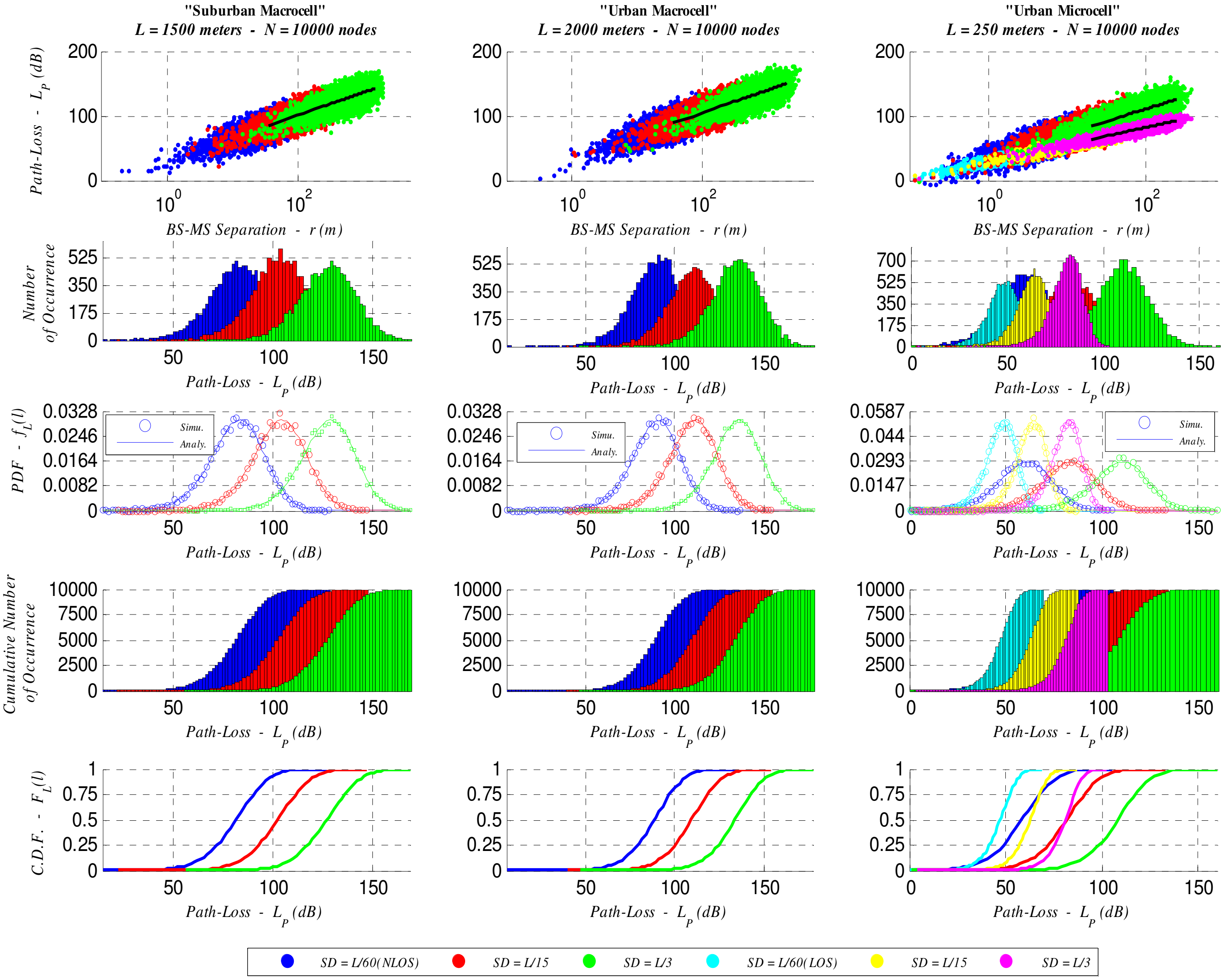

Figure 3. PL simulation based on IEEE 802.20 channel environments.

## V. Conclusion

The use of the Gaussian spatial distribution model is customary during the analysis and design of cellular systems, as well as air deployed sensor networks. However, to the best of our knowledge, though because of simplicity it has been shown for the uniform case, no channel loss for a Gaussianly spread network is available in literature. Therefore, in this paper, we for the first time obtained exact closed-form stochastic notation for this objective. In fact, we did this not only for mathematical elegance, but because it is more efficient for analysis as oppose to Monte Carlo simulation, the mathematical dependencies are more evident during examination, and it could be practical for reusability in other derivations of relevance to power consumption, coverage, detection, and sensing capability.

We also verified the analytical derivation through simulation, and as expected the PL distribution appropriately matches. Moreover, because the channel loss density model was deliberately derived with generic parameters, it is hence practical and can be applied for any cellular technology or WSN during the design phase by network engineers. Nonetheless, to demonstrate the analysis, we have based our simulation using specifications from the new IEEE 802.20, 4G mobile protocol. In future endeavors, it would be interesting to further expand this topic by attempting to find a channel predictor for the multihop/ad hoc network scheme.